\documentclass[12pt,a4paper,english,superscriptaddress,aps,prd,preprint,showpacs]{revtex4-1}
\usepackage{amssymb}
\usepackage{amsmath}
\usepackage{graphicx}
\usepackage{babel}
\usepackage{hyperref}
\usepackage{color}

\begin{document}

\title{Dynamical gauge symmetry breaking in lower-dimensional Lorentz-violating supersymmetric theory}

\author{A.~C.~Lehum}
\email{lehum@ufpa.br}
\affiliation{Faculdade de F\'isica, Universidade Federal do Par\'a, 66075-110, Bel\'em, Par\'a, Brazil.}

\author{T. Mariz} \email{tmariz@fis.ufal.br} \affiliation{Instituto de F\'{i}sica, Universidade Federal de Alagoas, 57072-900, Macei\'{o}, Alagoas, Brazil}

\author{J. R. Nascimento}
\affiliation{Departamento de F\'{\i}sica, Universidade Federal da Para\'{\i}ba,
 Caixa Postal 5008, 58051-970, Jo\~ao Pessoa, Para\'{\i}ba, Brazil}
\email{jroberto, petrov@fisica.ufpb.br}

\author{A. Yu. Petrov}
\affiliation{Departamento de F\'{\i}sica, Universidade Federal da Para\'{\i}ba,
 Caixa Postal 5008, 58051-970, Jo\~ao Pessoa, Para\'{\i}ba, Brazil}
\email{jroberto, petrov@fisica.ufpb.br}

\begin{abstract}
In this work we study the dynamical generation of mass in the Lorentz-violating low-dimensional Super-Yang-Mills theory in the aether superspace coupled to a scalar matter. We also suggest that our studies can be applied for condensed matter systems, especially lower-dimensional superconductors and topological insulators. In low dimensional materials, the parameter $\Delta$ generated by presence of the aether term can be interpreted as a quantity that renormalizes the propagation velocity of the bosonic mode with respect to the Fermi velocity. 
\end{abstract}

\pacs{11.30.Pb, 11.30.Cp, 11.15.Ex} 

\maketitle

\section{Introduction}\label{intro}

Modern studies of condensed matter call major attention to lower-dimensional quantum systems. The paradigmatic example is graphene representing itself as the most known lower-dimensional media. Within this context, application of quantum field theory methods is of special importance. For example, in \cite{lowenergy} it was claimed that the supersymmetry can emerge in low-energy physics of quantum materials such as superconductors and Weyl semimetals, within the context of (2+1)-dimensional and (1+1)-dimensional models. Clearly, it calls interest to low-dimensional supersymmetric models, especially, to supersymmetric Yukawa models. Many earlier, a possibility to apply supersymmetry within condensed matter study has been discussed also in \cite{Araujo}. It is clear that, to describe motion of charged particles within condensed matter, one should develop a supersymmetric theory involving gauge fields as well, while in three-dimensional space-time, the Chern-Simons term could play an important role. Some discussion of supersymmetry within the topological isolators context, together with a prescription for a possible experimental studies of such systems, is presented also in \cite{Ponte}.

At the same time, within condensed matter, the Lorentz symmetry breaking emerges naturally -- either in the case of presence of magnetic field, or due to a natural anisotropy of crystals. In four-dimensional models, Lorentz symmetry breaking is usually introduced in condensed matter through adding the Carroll-Field-Jackiw (CFJ) term (or, as is effectively the same, the axion term), as it is done in \cite{Grushin,Cambiaso}. However, in three-dimensional case the Chern-Smons term, representing itself as an analog of the CFJ term, is Lorentz invariant, and we need other terms to implement the Lorentz symmetry breaking. Apparently, the simplest Lorentz-breaking term in $3D$ (as well as in $2D$) is the aether term \cite{aether}. It is worth to mention that the aether term in scalar-spinor theory with Yukawa coupling, but without supersymmetry yet, has been considered within the context of $2D$ materials in \cite{Correa}. Therefore, it is natural to develop a supersymmetric theory involving it as an ingredient, with an intention to apply it further within the condensed matter context.

To do this, we follow the simplest way of applying the aether superspace which is a construction of a supersymmetric field theory allowing the use the powerful techniques of the superfield formalism, by the deformation of the supersymmetry (SUSY) algebra for supergauge field theories based on the Kostelecky-Berger construction \cite{KostBer}. 

As discussed before, the extension of the usual superspace to a two~\cite{Lehum:2015dqr} and three-dimensional~\cite{Farias:2012ed,Lehum:2013pca,Ganai} aether superspace is stated through the deformation of the SUSY generators
\begin{eqnarray}
Q_{\alpha}&=&i[\partial_{\alpha}-i\theta^{\beta}\gamma^m_{\beta\alpha}(\partial_m+k_{mn}\partial^n)]\nonumber\\
&=&i[\partial_{\alpha}-i{\theta}^{{\beta}}\gamma^m_{\beta\alpha}\nabla_m],
\end{eqnarray}
\noindent which satisfy the anti-commutation relation
\begin{eqnarray}
\{Q_{\alpha},Q_{\beta}\}=2i\gamma^m_{{\alpha}\beta}\nabla_m,
\end{eqnarray}
\noindent where $\nabla_m=\partial_m+k_{mn}\partial^n$, and $\partial_{\alpha}$ is the derivative with respect to the Grassmannian coordinates $\theta^{\alpha}$. We use Latin letters to denote indices of the space-time coordinates ($0,1$ for 2D and $0,1,2$ for 3D) and Greek letters for spinorial indices. The tensor $k_{mn}$ is a constant tensor which in the simplest case can be chosen to assume an aether-like form $k_{mn}=\alpha u_mu_n$, with $u^m$ being a constant vector (cf. \cite{aether}), with its square is either $-1$, 0 or 1, and $|\alpha|\ll 1$. We can also consider a more interesting case, where $k_{mn}$ is a traceless tensor, i.e., $k_{mn}=u_mu_n-\frac{1}{D}g_{mn}u^2$, with $D$ being the space-time dimension (for details, see \cite{Mariz:2016ooa}).


The aether-supercovariant derivative consistent with the deformed supersymmetry must anti-commute with SUSY generators $Q_\alpha$. It is given by
\begin{eqnarray}
\label{sder3d}
D_{\alpha}=\partial_{\alpha}+i{\theta}^{{\beta}}\gamma^m_{{\beta}\alpha}\nabla_m~,
\end{eqnarray}
where the derivative $\nabla_m$ commutes with the SUSY generators as well as with the supercovariant derivative $D_{\alpha}$.

In this paper, we consider various aspects of the lower (two and three)-dimensional super-Yang-Mills theory, especially, the quantum corrections in this theory. In the section 2, we perform quantum calculations aimed to generate one-loop contributions to the effective action. And the section 3 is our Summary where the results are discussed.

\section{The low-dimensional Super-Yang-Mills theory} \label{sym-2d}

As discussed in Ref.~\cite{Gomes:2011aa}, there is no substantial difference between conventions and notations for supersymmetric models defined in three- and two-dimensional space-time. Therefore we use the notations and conventions as in Ref.~\cite{SGRS}.  Our starting point is the classical action of the two and three-dimensional $SU(N)$ Super-Yang-Mills theory coupled to matter superfields defined in the aether superspace,
\begin{eqnarray}\label{eq1}
S=Tr\int{d^nxd^2\theta}\Big{\{}&&\frac{1}{2}W^{\alpha}W_{\alpha}
-\frac{1}{4\xi} D^{\alpha}\Gamma_{\alpha}D^2D^{\beta}\Gamma_{\beta}+\dfrac{1}{2}\bar{c}D^{\alpha}\left(D_{\alpha}c-ie[\Gamma_\alpha,c]\right)\nonumber\\
&&-\bar\Phi (D^2+m)\Phi-\frac{g^2}{2}\bar\Phi\Gamma^{\alpha}\Gamma_{\alpha}\Phi
+i\frac{g}{2}\left(D^{\alpha}\bar{\Phi}\Gamma_{\alpha}\Phi -\bar\Phi \Gamma^{\alpha}D_{\alpha}\Phi\right)\Big{\}},
\end{eqnarray}

\noindent where $W^{\alpha}=\frac{1}{2}D^{\beta}D^{\alpha}\Gamma_{\beta}-\dfrac{ig}{2}[\Gamma^{\beta},D_{\beta}\Gamma_{\alpha}]-\dfrac{g^2}{6}[\Gamma^{\beta},\{ \Gamma_{\beta},\Gamma_{\alpha}\}]$ is the gauge aether-superfield strength which transforms covariantly, $W'_{\alpha}=e^{iK}W_{\alpha}e^{iK}$, with $K=K(x,\theta)$ being a real scalar aether-superfield. The $n$ stands for the dimension of spacetime, allowing to assume $2$ or $3$. Within this paper, we assume that all fields are Lie algebra-valued, $\Gamma^{\alpha}=\Gamma^{\alpha a}T^a$, $\Phi=\Phi^aT^a$, etc., with $T^a$ being the gauge group generators. And the trace is assumed to be the trace of products of these generators. 

In order to obtain the gauge aether superfield propagator, it is convenient to write the quadratic part of the gauge aether superfield action as
\begin{eqnarray}\label{eq1a}
S_2&=&Tr\int{d^nxd^2\theta}\Big{\{}-\frac{1}{8}\Gamma_{\gamma}D^{\alpha}D^{\gamma}D^{\beta}D_{\alpha}\Gamma_{\beta}
-\frac{1}{4\xi} \Gamma_{\alpha}D^{\alpha}D^2D^{\beta}\Gamma_{\beta}\Big{\}}\nonumber\\
&=&Tr\int{\frac{d^np}{(2\pi)^2}}d^2\theta\Big{\{}-\frac{1}{4}\Gamma_{\gamma}(\tilde{p},\theta)\tilde{p}^2\left(C_{\beta\gamma}+\frac{\tilde{p}_{\beta\gamma}D^2}{\tilde{p}^2}\right)\Gamma_{\beta}(-\tilde{p},\theta)-\frac{1}{4\xi} \Gamma_{\alpha}D^{\alpha}D^2D^{\beta}\Gamma_{\beta}
\Big{\}},
\end{eqnarray}
\noindent where {$C^{\alpha\beta}=i\epsilon^{\alpha\beta}$ is an Hermitian antisymmetric matrix} $\tilde{p}_{\beta\gamma}=(\gamma^m)_{\beta\gamma}\tilde{p}_m=(\gamma^m)_{\beta\gamma}(p_m+k_{mn}p^n)$ is the twisted moment, $\tilde{p}^2=p^2+2k_{mn}p^mp^n+k^{mn}k_{ml}p_np^l$ and $D^2=\partial^2-\theta^\beta(\gamma^m)_{\beta\alpha}\tilde{p}_m\partial^\alpha +\theta^2\tilde{p}^2$.

The propagators obtained from Eq. (\ref{eq1}), can be cast as
\begin{eqnarray}\label{eq4}
\langle \Gamma^{\alpha}_a(-\tilde{p},\theta_1)\Gamma^{\beta}_b(\tilde{p},\theta_2)\rangle&=&\frac{i\delta_{ab}}{2}\frac{D^2}{(\tilde{p}^2)^2}
\left(D_{\beta}D_{\alpha}-\xi D_{\alpha}D_{\beta}\right)\delta_{12}\nonumber\\
&=&\frac{i\delta_{ab}}{2}\frac{(1+\xi)C_{\beta\alpha}\tilde{p}^2+(1-\xi)\tilde{p}_{\beta\alpha}D^2}{(\tilde{p}^2)^2}
\delta_{12}~,\label{propgauge}\\
\langle c_a(\tilde{p},\theta_1)\bar{c}_b(-\tilde{p},\theta_2)\rangle&=&i\delta_{ab}\frac{D^2}{\tilde{p}^2}\delta_{12}~,\nonumber\\
\langle \Phi_a(\tilde{p},\theta_1)\bar\Phi_b(-\tilde{p},\theta_2)\rangle&=&-i\delta_{ab}\frac{D^2-m}{\tilde{p}^2+m^2}\delta_{12}~,
\end{eqnarray}

\noindent
where $\delta_{12}=\delta^2(\theta_1-\theta_2)$, it is a well known Grassmannian delta function \cite{SGRS}. Without loss of generality, we choose to work in the Feynman gauge, i.e. $\xi=1$. We note that it is natural to suggest $|k_{mn}|\ll 1$ since the Lorentz symmetry breaking is small. Under this assumption, no ghosts or tachyons arise, and there is no any problems with stability (for discussion of dispersion relations in this theory, see also \cite{Farias:2012ed}).

\subsection{Pure gauge sector}

The effective action receives one-loop pure gauge sector contributions from the diagrams drawn in Figs. \ref{fig1} (a), (b) and (c). Performing the D-algebra manipulations with the help of the computer package SusyMath~\cite{Ferrari:2007sc}, we get the following results. Here we suggest that the gauge generators obey relations ${\rm tr}(T^aT^b)=\delta^{ab}$. The supergraph Fig. \ref{fig1}(a) is vanishing, while other contributions can be cast as
\begin{eqnarray}\label{eq5}
S_{\ref{fig1}(b)}=&&-N\frac{g^2}{4}\int{\frac{d^np}{(2\pi)^2}}d^2\theta~\int{\frac{d^nq}{(2\pi)^2}}~\Gamma_a^{\alpha}(\tilde{p},\theta)\frac{\left(\tilde{p}_{\alpha\beta}D^2+2C_{\beta\alpha}\tilde{q}^2\right)}{\tilde{q}^2(\tilde{q}+\tilde{p})^2}\Gamma_a^{\beta}(-\tilde{p},\theta);
\end{eqnarray}
\begin{eqnarray}\label{eq6}
S_{\ref{fig1}(c)}=&&-N\frac{g^2}{4}\int{\frac{d^np}{(2\pi)^2}}d^2\theta~\int{\frac{d^nq}{(2\pi)^2}}~\Gamma_a^{\alpha}(\tilde{p},\theta)\frac{\left(\tilde{p}^2-2\tilde{q}^2\right)C_{\beta\alpha}}{\tilde{q}^2(\tilde{q}+\tilde{p})^2}\Gamma_a^{\beta}(-\tilde{p},\theta).
\end{eqnarray}

Adding the two diagrams above, after some algebraic manipulations, we have
\begin{eqnarray}\label{eq7}
S_{1loop}&=&-N\frac{g^2}{4}\int{\frac{d^np}{(2\pi)^2}}d^2\theta~\Gamma_a^{\gamma}(\tilde{p},\theta)\left(\tilde{p}_{\alpha\beta}D^2+C_{\beta\alpha}\tilde{p}^2\right)\Gamma_a^{\beta}(-\tilde{p},\theta)~\int{\frac{d^nq}{(2\pi)^2}}\frac{1}{\tilde{q}^2(\tilde{q}+\tilde{p})^2}\nonumber\\
&=&-\frac{1}{4}\int{\frac{d^np}{(2\pi)^2}}d^2\theta~\Gamma_a^{\alpha}(\tilde{p},\theta)\left(C_{\beta\gamma}+\frac{\tilde{p}_{\beta\gamma}D^2}{\tilde{p}^2}\right)\tilde{p}^2~\Gamma_a^{\beta}(-\tilde{p},\theta)~\int{\frac{d^nq}{(2\pi)^2}}\frac{Ng^2}{\tilde{q}^2(\tilde{q}+\tilde{p})^2}.
 \end{eqnarray}
 
Summing up the one-loop correction Eq.(\ref{eq7}) to the classical part of the effective action Eq.(\ref{eq1a}), the pure gauge sector contributions to the effective action can be cast as
\begin{eqnarray}\label{eq70a}
S_{gauge}&=&-\frac{1}{4}\int{\frac{d^np}{(2\pi)^2}}d^2\theta~\Gamma_a^{\gamma}(\tilde{p},\theta)\left(C_{\beta\gamma}+\frac{\tilde{p}_{\beta\gamma}D^2}{\tilde{p}^2}\right)\tilde{p}^2\times\nonumber\\&\times&
\left[1+\int{\frac{d^nq}{(2\pi)^2}}\frac{Ng^2}{\tilde{q}^2(\tilde{q}+\tilde{p})^2}\right]\Gamma_a^{\beta}(-\tilde{p},\theta).
\end{eqnarray}
We note that since $\left(C_{\beta\gamma}+\frac{\tilde{p}_{\beta\gamma}D^2}{\tilde{p}^2}\right)\tilde{p}^2=D_{\beta}D_{\gamma}D^2$, this action is proportional to $\Gamma_a^{\gamma}D_{\beta}D_{\gamma}D^2\Gamma_a^{\beta}$, thus being perfectly gauge invariant.

In the $D=3$, the one-loop contribution to the above effective action is just a non-local correction to the $-\frac{1}{8}\Gamma_{\gamma}D^{\alpha}D^{\gamma}D^{\beta}D_{\alpha}\Gamma_{\beta}$ term (that is, the Maxwell term). But, in two dimensions the situation is different.

To $D=2$, the integral over $k$ gives 
\begin{eqnarray}\label{eq7a}
S_{gauge}&=&-\frac{1}{4}\int{\frac{d^2p}{(2\pi)^2}}d^2\theta~\Gamma_a^{\gamma}(\tilde{p},\theta) \left(C_{\beta\gamma}+\frac{\tilde{p}_{\beta\gamma}D^2}{\tilde{p}^2}\right)\tilde{p}^2\left[1+\frac{Ng^2}{4\pi \tilde{p}^2}\right] \Gamma_a^{\beta}(-\tilde{p},\theta)\nonumber\\
&=&-\frac{1}{4}\int{\frac{d^2p}{(2\pi)^2}}d^2\theta~\Gamma_a^{\gamma}(\tilde{p},\theta)\left(C_{\beta\gamma}+\frac{\tilde{p}_{\beta\gamma}D^2}{\tilde{p}^2}\right)\left[\tilde{p}^2+\tilde{M}^2\right] \Gamma_a^{\beta}(-\tilde{p},\theta)
\end{eqnarray}

\noindent where $\tilde{M}^2=\Delta~M^2=\Delta~Ng^2/4\pi$. 
In the usual notations, we can rewrite this expression as
\begin{eqnarray}
S_{gauge}&=&\frac{1}{2}\int{\frac{d^2p}{(2\pi)^2}}d^2\theta~W_a^{\alpha}(\tilde{p},\theta)
\left[1+\frac{\tilde{M}^2}{\tilde{p}^2}\right] W_{a\alpha}(-\tilde{p},\theta),
\end{eqnarray}
that is, the nonlocal extension of the Maxwell action. However, it is easy to check that this nonlocality essentially affects only the longitudinal sector, while in the transversal one, or as is the same, under the condition $D^{\alpha}\Gamma^a_{\alpha}=0$, it completely goes away. Here,
in the integral over $q$ we have changed the variable of integration $q$ to $\tilde{q}$, so that we can write $\int{d^3q}=\Delta\int{d^3\tilde{q}}$, where $\Delta=\det^{-1}(\delta^m_n+u_n u^m)$ is the Jacobian of the transformation. In low dimensional materials, $\Delta$ can be interpreted as a quantity that renormalizes the propagation velocity of the bosonic mode with respect to the Fermi velocity~\cite{Correa:2017rnm}.

For small $u_m$, we have $\Delta\approx(1-u^2)$. Alternatively, we can consider $k_{mn}$ to be traceless, i.e., $k_{mn}=u_mu_n-\frac{1}{D}g_{mn}u^2$ as above, where now $\Delta=\det^{-1}(\delta^m_n+\frac{3}{4}u_n u^m)\det^{-3}(\delta^m_n-\frac{1}{4}u_n u^m)$. In this case, for the small $u_m$, one evidently has $\Delta=1$. Moreover, in principle nobody forbids to consider an antisymmetric $k_{mn}$ which evidently yields $\Delta=1$ as well, i.e., there is no modification of the measure.

Even though the pure gauge sector of the model is massless at classical level, the charge $g$ is a dimensionful parameter with mass dimension one. Therefore, the pure aether-SYM$_{2}$ exhibit a dynamical generation of mass, where $\tilde{M}$ is a parameter dependent on the Lorentz breaking properties of the aether-superspace. 

In order to understand the character of the aether dependence on the gauge superfield mass, let us discuss with some detail the dispersion relation of the gauge superfield. From Eq.(\ref{eq7a}), it is easy to see that the gauge superfield propagator has a massive pole given by
\begin{equation}
\tilde{p}^{2}+\tilde{M}^{2}=p^{2}+2k_{mn}p^{m}p^{n}+k^{mn}k_{ml}p_{n}p^{l}+\tilde{M}^{2}=0\thinspace,
\end{equation}
where, in terms of the aether LV vector $u^{m}$, can be cast as
\begin{equation}
p^{2}+2\alpha\left[1+\alpha u^{2}\right]\left(u^{m}p_{m}\right)^{2}+\tilde{M}^{2}=0\thinspace.\label{eq:DR}
\end{equation} 
The form of this dispersion relation is similar to that one from  \cite{Farias:2012ed,Lehum:2013pca}. Here, we give the detailed analysis of this relation for various situations.

First, we are able to analyze the consequences of Eq.\,\eqref{eq:DR} for $u^{m}u_{m}$ being $\pm1$ or zero. Let us start  with spacelike $u^{m}$, i.e., $u^{m}u_{m}=+1$, choosing coordinates such that $u^{m}=\left(0,\hat{u}\right)$, where $\hat{u}$ is a unitary space vector. With this choice, Eq.\,\eqref{eq:DR}
can be cast as
\begin{equation}
E^{2}={\vec{p}}^{2}+M^{2}(1-\alpha)+\alpha(2+\alpha)(\hat{u}\cdot\vec{p})^2\thinspace.
\end{equation}

In the rest frame, $\vec{p}=\vec{0}$, we find $E^2=M^2(1-\alpha)$, resulting in a LV background dependence to the  mass of the particle.

If $u^{m}$ is a timelike vector, i.e., $u^{m}u_{m}=-1$, we can choose $u^{m}=\left(1,\vec{0}\right)$. Thus, the dispersion relation can be written as
\begin{equation}
E^{2}=\frac{\vec{p}^{2}+M^{2}(1+\alpha)}{1+2\alpha(1-\alpha)}\thinspace.
\end{equation}

In the rest frame, we find
\begin{equation}
E^{2}=\frac{M^{2}(1+\alpha)}{1+2\alpha(1-\alpha)}\approx M^2(1-\alpha)+\mathcal{O}(\alpha^2)\thinspace.
\end{equation}
Notice that the generated mass for $u^{m}$ spacelike and timelike vector is the same, up to $\mathcal{O}(\alpha)$.

Finally, for the lightlike case, we have $u^2=0$ and $\Delta=1$. Therefore, the dispersion relation can be cast as
\begin{equation}
E^{2}\left[1-2\alpha\left(u^{0}\right)^{2}\right]+4\alpha u^{0}\left(\vec{u}\cdot\vec{p}\right)E-2\alpha\left(\vec{u}\cdot\vec{p}\right)^{2}-\vec{p}^{2}-\tilde{M}^{2}=0\thinspace.
\end{equation}

Let us chose the reference frame such that $u^m=\left(u^{0},u^{0},0,0\right)$. Let us consider two different situations. First, let $\vec{p}$ parallel to $\vec{u}$. In this case, the dispersion relation becomes
\begin{equation}
E=\frac{1}{1-2\alpha}\left[-2\alpha|\vec{p}|\pm\sqrt{\vec{p}^{2}+M^{2}\left(1-2\alpha\right)}\right]\thinspace.
\end{equation}
If we set $\vec{p}$ perpendicular to $\vec{u}$, we
obtain
\begin{equation}
E^{2}=\frac{\vec{p}^{2}+M^{2}}{1-2\alpha}\thinspace.
\end{equation}
In both cases, in the rest frame, we find the following dispersion relation up to order $\mathcal{O}(\alpha)$
\begin{equation}
E^{2}=M^2(1+2\alpha)\thinspace.
\end{equation}

It is important to note that in the non-abelian gauge theory considered here, the dynamical generated mass is dependent on the aether properties in every possible situation. The massive pole of gauge superfield is $M^2(1-\alpha)$ for spacelike and timelike vector $u^m$, and $M^2(1+2\alpha)$ for lightlike vector $u^m$. It is different from what happens in another gauge theory, $CP^{(N-1)}$ model, where in some special cases the dynamical generated mass is independent on the aether parameters \cite{Ferrari:2017rwk}.
 
\subsection{Matter couplings}

\begin{figure}[ht]
	\includegraphics[angle=0,width=14cm]{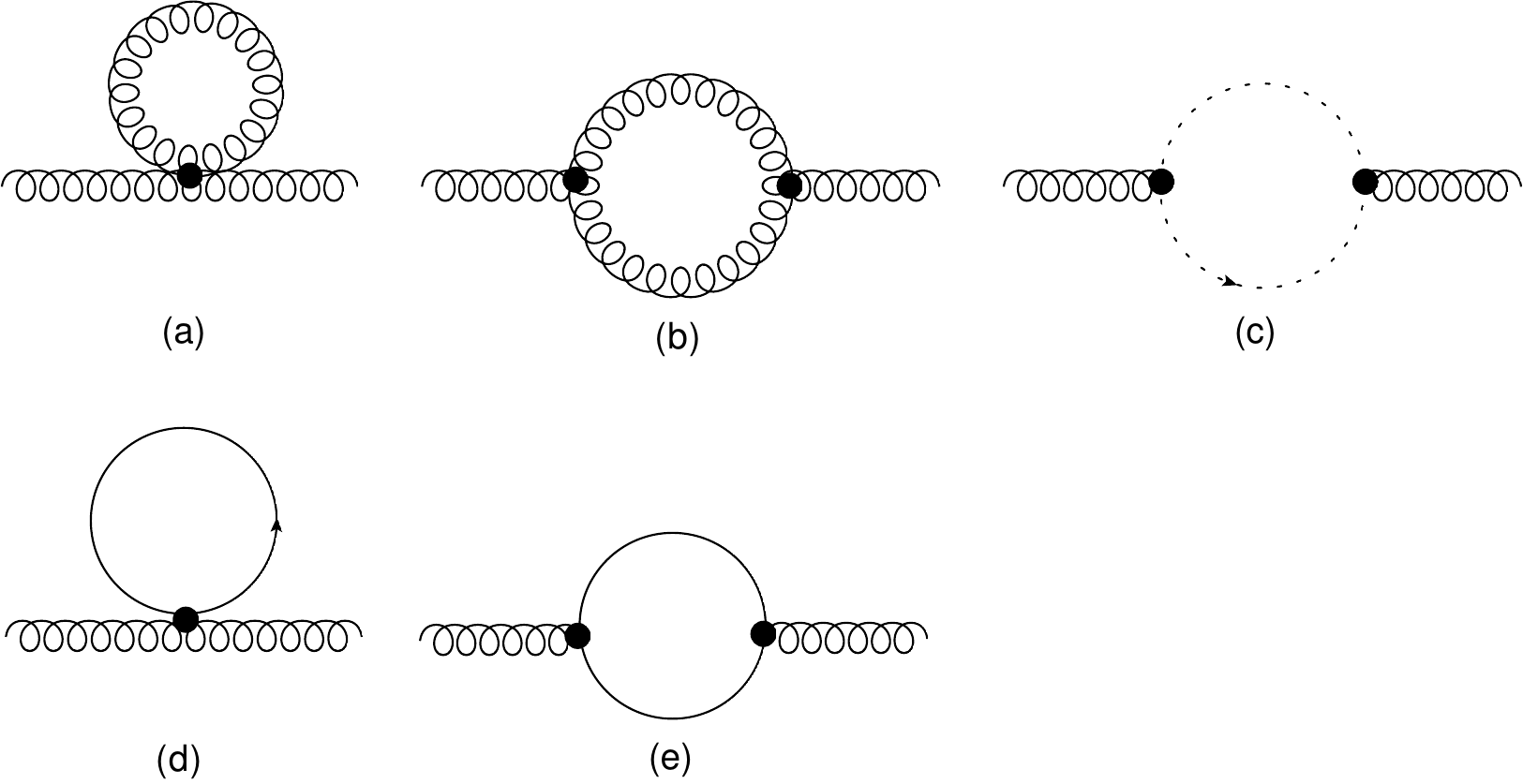}
	\caption{One-loop contributions to the gauge superfield effective action. Wiggly, dashed and continuous lines represent the gauge, ghost and matter superfield propagators, respectively.}
	\label{fig1}
\end{figure}

Now, let us compute the one-loop corrections to the effective action of the gauge aether-superfield due to matter interactions, Figs.~\ref{fig1}(d) and \ref{fig1}(e). The contributions Fig.~\ref{fig1}d and Fig.~\ref{fig1}e are given by
\begin{equation}\label{eq001}
S_{m\ref{fig1}a} =-g^2\frac{N}{2}\int\frac{d^np}{(2\pi)^2}d^2\theta
~{\Gamma_a}_{\beta}(-\tilde{p},\theta)\int\frac{d^nq}{(2\pi)^2}\frac{C^{\alpha\beta} }{\tilde{q}^2+m^2}{\Gamma_a}_{\alpha}(\tilde{p},\theta),
\end{equation}

\noindent and 
\begin{eqnarray}\label{eq002}
S_{m\ref{fig1}b}  & = &g^2 \frac{N}{2}\int\frac{d^np}{(2\pi)^2}d^2\theta~{\Gamma_a}_{\beta}(-\tilde{p},\theta) \int\frac{d^nq}{(2\pi)^2}\frac{1}{[(\tilde{q}+\tilde{p})^2+m^2](\tilde{q}^2+m^2)}\nonumber\\
&\times&\left[(\tilde{q}^2+m^2)C^{\alpha\beta}+(\tilde{q}^{\alpha\beta}+mC^{\alpha\beta})D^2+
\frac{1}{2}(\tilde{q}^{\gamma\beta} +mC^{\gamma\beta})D_{\gamma}D^{\alpha}\right] {\Gamma_a}_{\alpha}(\tilde{p},\theta),
\end{eqnarray}

\noindent respectively. Adding the two contributions above, we have
\begin{eqnarray}\label{eq003}
S_{m}& = &g^2\frac{N}{2}\int\frac{d^np}{(2\pi)^2}d^2\theta~{\Gamma_a}^{\beta}(-\tilde{p},\theta) \int\frac{d^nq}{(2\pi)^2}\frac{\tilde{q}_{\beta\gamma}-mC_{\beta\gamma}}{[(\tilde{q}+\tilde{p})^2+m^2](\tilde{q}^2+m^2)}{\tilde{W}_a}^\gamma(\tilde{p},\theta).
\end{eqnarray}

\noindent where ${\tilde{W}_a}^{\alpha}=\frac{1}{2}D^{\beta}D^{\alpha}{\Gamma_a}_{\beta}$ is the linear part of the Yang-Mills aether-superfield strength $W^{\alpha}$.

Through the identity
\begin{eqnarray}\label{eq004}
&&\int\frac{d^nq}{(2\pi)^2}\frac{\tilde{q}^{\alpha\beta}}{[(\tilde{q}+\tilde{p})^2+m^2](\tilde{q}^2+m^2)}\nonumber\\&=& 
-\frac{\tilde{p}^{\alpha\beta}}{2}\int\frac{d^nq}{(2\pi)^2}\frac{1}{[(\tilde{q}+\tilde{p})^2+m^2](\tilde{q}^2+m^2)} =
-\frac{\tilde{p}^{\alpha\beta}}{2}f(\tilde{p}),
\end{eqnarray}
the Eq.(\ref{eq003}) can be rewritten as 
\begin{eqnarray}\label{eq005}
S_{m} & = &-g^2\frac{N}{2} \int\frac{d^np}{(2\pi)^2}d^2\theta~ f(\tilde{p})\left[{\tilde{W}_a}^{\alpha}(-\tilde{p},\theta){\tilde{W_a}}_{\alpha}(\tilde{p},\theta)+2m\Gamma_a^{\alpha}(-\tilde{p},\theta){\tilde{W_a}}_{\alpha}(\tilde{p},\theta)\right].
\end{eqnarray}
It is interesting to consider the low-energy limit of this expression, that is, to keep only the leading terms at $p\to 0$. In this case one has $f(\tilde{p})|_{p\to 0}=\frac{\Delta}{8\pi|m|}$. So, our result takes the form
\begin{eqnarray}\label{eq006}
S_{m} & \simeq &-g^2\frac{N}{2}\frac{\Delta}{8\pi|m|} \int\frac{d^np}{(2\pi)^2}d^2\theta~\left[{\tilde{W}_a}^{\alpha}(-\tilde{p},\theta){\tilde{W_a}}_{\alpha}(\tilde{p},\theta)+2m\Gamma_a^{\alpha}(-\tilde{p},\theta){\tilde{W_a}}_{\alpha}(\tilde{p},\theta)\right].
\end{eqnarray}
We conclude that we succeeded to generate the aether-like supersymmetric Maxwell-Chern-Simons term.

It is well-known that the presence of Chern-Simons (CS) term $\Gamma^\alpha {W_0}_\alpha$ generates a topological massive pole to the gauge aether superfield propagator. In two dimensions, the above expression represents the effective action for a massive gauge invariant aether superfield, just as discussed in Refs. \cite{Gates:1977hb,Bengtsson:1983in} in a two-dimensional ordinary superspace. Note nevertheless that the massive term vanishes in two dimensions.


\section{Final remarks}\label{remarks}

In this work we studied the perturbative generation of the linearized Yang-Mills and Chern-Simons terms in the aether superspace. We explicitly demonstrated that this generation requires no more difficult calculations than in the usual superspace. Unlike the earlier studies \cite{Lehum:2015dqr,Lehum:2013pca}, we, for the first time, considered the contributions from essentially non-Abelian vertices. We explicitly showed that the nonlocality present in the term arising from the purely gauge sector is nonphysical as it does not affect the transversal part of the gauge superfield. Finally, we argued that the same approach can be applied to generation of the full-fledged non-Abelian super-Yang-Mills theory as well, the details will be presented in our next work. Also, we expect to apply our studies for condensed matter systems, especially lower-dimensional superconductors and topological insulators. In low dimensional materials, the aether parameter $\Delta$ can be interpreted as a quantity that renormalizes the propagation velocity of the bosonic mode with respect to the Fermi velocity~\cite{Correa:2017rnm}.

As another possible interpretation, we can also suggest to treat the Lorentz symmetry breaking as an effective description of polarization or magnetization of the media, therefore, the factor $\Delta$ arising within our calculations naturally plays the role of electric or magnetic permeability.

\acknowledgments{This work was partially supported by Conselho Nacional de Desenvolvimento Cient\'{\i}fico e Tecnol\'{o}gico (CNPq). The work by A. Yu. P. has been supported by the CNPq project 303783/2015-0. A. C. L. has been partially supported by the CNPq project 307723/2016-0 and 402096/2016-9.}

\vspace*{2mm}

\end{document}